\documentclass[aps,prl,reprint,twocolumn,showpacs,superscriptaddress,groupedaddress,footinbib,floatfix]{revtex4-2}

\usepackage{graphicx,natbib,amsmath}
\usepackage{dcolumn}
\usepackage{bm}
\usepackage[colorlinks=True,citecolor=blue,linkcolor=blue,urlcolor=blue]{hyperref}
\usepackage{xspace}
\usepackage{aas_macros}

\newcommand{\rgc}{$r_\mathrm{g}/c$\xspace}

\usepackage{verbatim}

\begin{document}


\title{Particle Acceleration in Collisionless Magnetically Arrested Disks}

\author{Jesse Vos$^{1}$}
\email{jt.vos@astro.ru.nl}
\author{Beno\^it Cerutti$^{2}$}
\author{Monika Mo\'scibrodzka$^{1}$}
\author{Kyle Parfrey$^{3}$}
\affiliation{
$^{1}$Department~of~Astrophysics/IMAPP,~Radboud~University,~PO~Box~9010,~6500~GL~Nijmegen,~the~Netherlands\\
$^{2}$Univ.~Grenoble~Alpes,~CNRS,~IPAG,~38000~Grenoble,~France\\
$^{3}$Princeton Plasma Physics Laboratory, Princeton, NJ 08540, USA
}


\date{\today}

\begin{abstract}
We present the first collisionless realization of two-dimensional axisymmetric black hole accretion consistent with a persistent magnetically arrested disk state.
The accretion flow, consisting of an ion-electron disk plasma combined with magnetospheric pair creation effects, is simulated using first-principles general-relativistic particle-in-cell methods. 
The simulation is evolved over significant dynamical timescales during which a quasi-steady accretion state is reached with several magnetic flux eruption cycles.
We include a realistic treatment of inverse Compton scattering and pair production, which allows for studying the interaction between the collisionless accretion flow and pair-loaded jet. 
Our findings indicate that magnetic flux eruptions associated with equatorial magnetic reconnection within the black hole magnetosphere and the formation of spark gaps are locations of maximal particle acceleration.
Flux eruptions, starting near the central black hole, can trigger Kelvin-Helmholtz-like vortices at the jet-disk interface that facilitate efficient mixing between disk and jet plasma in this region. 
Transient periods of increased pair production following magnetic flux eruptions and reconnection events are responsible for most of the highly accelerated particles. 
\end{abstract}

\maketitle

\paragraph{Introduction.---}
Supermassive black holes (BH) at the center of low-luminosity active galactic nuclei (LLAGN) accrete ion-lepton plasma whose high temperatures and low densities place it in the collisionless regime \footnote{The collisionless regime refers to the mean free path to Coulomb collisions being large -- often exceeding the system size by a multitude of factors \citep[e.g,][]{fitzpatrick14}}.
Some of these BHs are known to launch jets, but these features are well-established to not be universal \citep{madejski16,blandford19,hada20}.
Examples can be found in the two LLAGN M87$^\ast$ and Sagittarius A$^\ast$, where M87$^\ast$ shows a prominent, large-scale jet and Sagittarius A$^\ast$ only displays radiative characteristics typical for a compact jet \citep{markoff07,falcke09}.
The Event Horizon Telescope has observed emission structures associated with millimeter-wavelength synchrotron radiation from near-horizon relativistic accretion flows for both systems \citep{eht1,eht22sgrai}.
Although these emission features are likely produced by the bulk accretion flow, the question remains of how and where the higher-energy emission and particles are created.

Many jetted black hole systems, including M87$^\ast$ and Sagittarius A$^\ast$, display $\gamma$-ray activity \citep{aharonian06sgra,aharonian06m87,magic12m87,aliu12veritasm87,hess18sgra,magic20sgra,adams21veritassgra}.
These systems can go through periods of increased (i.e., flaring) activity \citep{eht21multiwavelengthm87,eht22sgraii,hess24m87}, which can be related to the dynamics of the central engine that drives jet launching mechanics and its associated plasma effects.
The source of the high-energy emission is often tied to the collisionless plasma dynamics within the BH magnetosphere, but it remains a challenge to identify the responsible mechanisms and locations.
With this numerical study, we therefore aim to simultaneously assess the ability of accretion-driven and magnetospheric processes to accelerate particles to high, non-thermal energies. 

The Magnetically Arrested Disk \citep[MAD;][]{igumenshchev03,narayan03} state is currently a preferred model for interpreting millimeter emission from the two aforementioned LLAGN, both with regard to the polarimetric signatures \cite{eht8m87pol21} and the origin of flares with associated orbital motion of hot spots \citep{gravity_s2_18,gravity_jimenez20,wielgus22b,vos22}. 
MAD episodes are characterized by flux eruptions, during which a limiting horizon-penetrating magnetic flux is reached, or even exceeded, resulting in the partial halting of the accretion flow \citep[e.g.,][]{tchekhovskoy11,mckinney12,ripperda20,ripperda22,davelaar23,vos24timelags}.
General-relativistic magnetohydrodynamical (GRMHD) models that bring about the MAD state are characterized by large-scale coherent magnetic fields and sizable initialized disk structures.
However, GRMHD methods are technically not applicable in the collisionless regime and do not directly provide insight into the non-ideal processes associated with particle acceleration.
It is therefore valuable to investigate these systems with methods that self-consistently take into account these processes while simultaneously building on the GRMHD findings.

In this Letter, we employ general relativistic (radiative) particle-in-cell (GR[R]PIC) methods to undertake a global collisionless study combining initialized accretion flows and particle creation processes. 
We present a quasi-steady, axisymmetric MAD accretion state and focus on interpreting the locations and mechanisms of particle acceleration within this environment. 
The system's characteristics --- highly magnetized, with large-scale coherent field structures --- are reminiscent of the well-established MAD state in GRMHD models 
\citep[e.g.,][]{tchekhovskoy11,mckinney12,ripperda20,ripperda22,davelaar23,vos24timelags}.
Previously, several kinetic studies of black hole magnetospheres were undertaken that mostly focused on pair-created plasma dynamics under magnetically dominated conditions \cite{parfrey19,crinquand20,crinquand21,crinquand22,elmellah22,elmellah23,niv23} or spherically-symmetric Bondi-Hoyle accretion \citep{galishnikova23}.
Here, we present the first realization of disk-like initial conditions in a collisionless scenario, consisting of a three-species (pair-ion) plasma with a significant ion-to-electron mass ratio.
With the addition of a strong accretion flow component to the BH magnetosphere in the presented simulation, it becomes possible to better constrain the flaring dynamics of AGN systems and gain insight into the edge-brightened mechanisms of AGN jets \citep{perucho07,kim18,janssen21,pasetto21,issaoun22,rieger22}.

\paragraph{Methods.---} 
We use the GRRPIC code {\tt Zeltron} \cite{cerutti13,parfrey19} to evolve an ion-electron accretion flow combined with a pair-filled black hole magnetosphere.
The total simulation domain is denoted by $r \in [0.9r_\mathrm{h}, 27r_\mathrm{g}]$ and $\theta \in [0.01\pi,0.99\pi]$ with a resolution of $N_r \times N_\theta = 1440^2$ for spherical Kerr-Schild coordinates with a grid that is logarithmically spaced in $\hat{r}$ and linearly spaced in $\hat{\theta}$.
Here, $r_\mathrm{g} = GM / c^2$ is the gravitational radius and $r_\mathrm{h} = (1 + \sqrt{1-a^2})r_\mathrm{g}$ is the horizon radius, with gravitational constant $G$, black hole mass $M$, speed of light $c$, and dimensionless black hole spin $a = 0.99$.
The ion mass is set to $m_i = 100\, m_e$, where $m_e$ is the electron mass.
We initialize a uniform number density, $n = 8\, n_\textrm{GJ}$, wedge that spans $r \in [5, 27]r_\mathrm{g}$, $\theta \in [\pi/3,2\pi/3]$, where $n_\mathrm{GJ} = B_0 \omega_\mathrm{h} / (4 \pi c e)$ is the Goldreich-Julian density with angular velocity $\omega_\mathrm{h} = a c / (2r_\mathrm{h})$ \citep{goldreich69} and $B_0$ the initial magnetic field strength \citep[see Sup.~Mat.;][]{supplemental}.
In the inner region, i.e., $r \in [r_\mathrm{h}, 5\, r_\mathrm{g}]$, we initialize a uniform pair-plasma sphere with a number density of $1 \, n_\mathrm{GJ}$ to kickstart pair production near the black hole.
The magnetic field begins as a single large magnetic loop in the poloidal plane \citep[see Sup.~Mat.;][]{supplemental} with an initialized ratio of thermal to magnetic pressure $\beta \in [0.1, 5]$.
Our primary focus is investigating the collisionless implications of a persistent MAD accretion state.  

The initial configuration of the ion-electron plasma is somewhat arbitrarily chosen and of little consequence to the formation of the relaxed, quasi-steady axisymmetric state that is achieved after the simulation has evolved for sufficient time \citep[see Sup.~Mat.;][]{supplemental}.
No angular momentum is introduced for the accretion flow at initialization, which expedites the onset of an axisymmetric quasi-steady accretion state.
However, we note that frame-dragging 
effects will quickly introduce angular momentum for the innermost black hole regions.
For simplicity, we refer to the low angular-momentum accretion flow as the ``disk".
The ion-electron disk plasma is initialized with a temperature $k_\mathrm{B} T_\mathrm{inj} = m_e c^2$ for both ions and electrons.
Plasma is replenished from a spherical shell corresponding to $r \in [25, 27]r_\mathrm{g}$, $\theta \in [\pi/3,2\pi/3]$, where injection occurs at a fixed rate when the density drops below the initialized disk density.
The injection from the boundary emulates the inflow of matter from a large-scale accretion flow.
Particles that pass through the outer boundary are removed \citep{cerutti15}.

Electron-positron pairs are created via photon-photon annihilation of inverse-Compton scattered $\gamma$-ray photons with a background photon field \citep{supplemental}. 
The inclusion of these particle creation effects provides insight into how and if the jet is loaded with pairs and whether these pairs are able to significantly interact at the jet-disk interface.
We refer the reader to the Supplemental Material \citep{supplemental} (and references \citep{levinson18,crinquand20}) for a more detailed description of the simulation specifics, such as the definition of the poloidal magnetic field initialization, pair production criteria, and characteristic scaling.

\begin{figure}
    \centering
    \includegraphics[width=\linewidth]{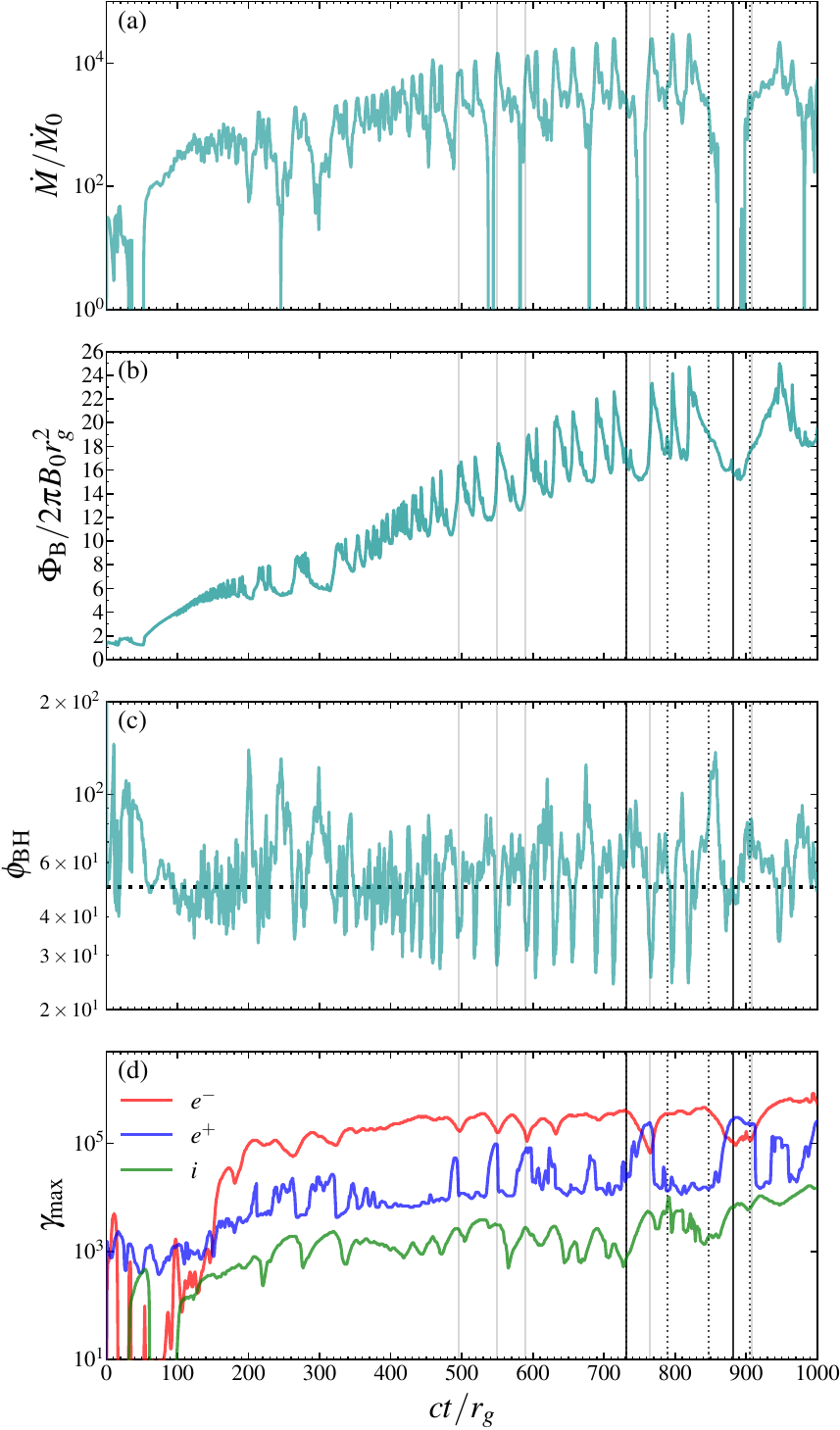}
    \caption{
    Time series of the integrated horizon-penetrating fluxes and acceleration of particles.
    The panels display the (a) combined ($i$, $e^-$, $e^+$) mass accretion rate ($\dot{M}$), (b) horizon-penetrating magnetic flux ($\Phi_\mathrm{B}$), (c) normalized magnetic flux known as the MAD parameter ($\phi_\mathrm{BH}=\Phi_\mathrm{B}/\sqrt{\dot{M}r^2_\mathrm{g} c}$), and (d) the maximum Lorentz factor ($\gamma_\mathrm{max}$, see also Fig.~\ref{fig:spectra}) for the three particle species.
    The horizontal line in panel (c) denotes the theoretically established limit of $\phi_\mathrm{BH} \sim 50$. The mass accretion rate is normalized with a fiducial quantity defined as $\dot{M}_0 = 0.3 c \, m_i \, n_\mathrm{GJ} \, 2 \pi r_\mathrm{h}^2$.
    The vertical gray lines mark several times when $\gamma_\mathrm{max, e^+} \gtrsim \gamma_\mathrm{max, e^-}$, while the vertical black lines indicate the times displayed in Fig.~\ref{fig:densmaps} (solid) and \ref{fig:mixing} (dotted).
    }
    \label{fig:timeseries}
\end{figure}

\begin{figure}
    \centering
    \includegraphics[width=\linewidth]{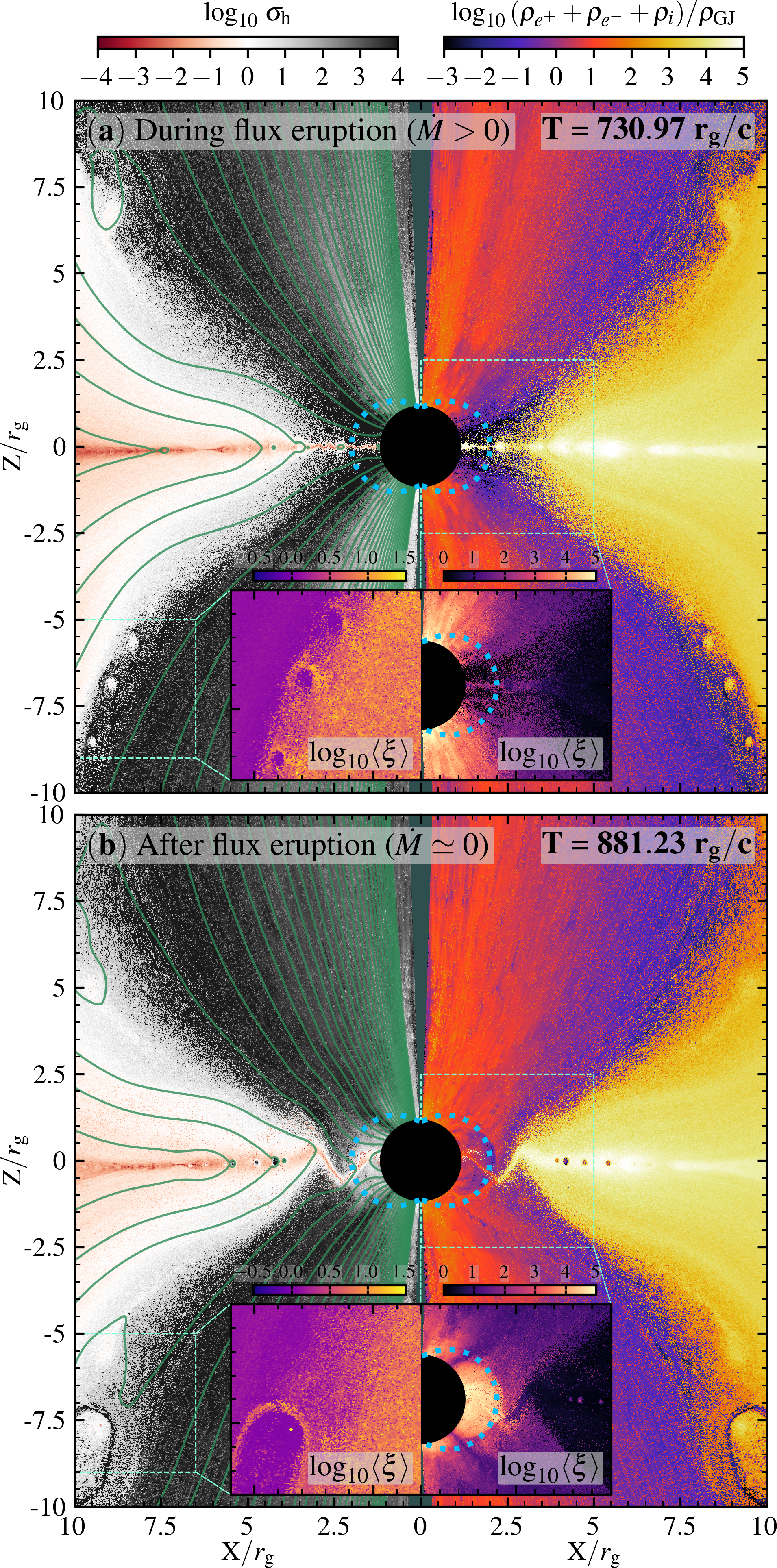}
    \caption{
    Overview of the axisymmetric MAD accretion state during and after a flux eruption. 
    The displayed quantities are magnetization $\sigma_\mathrm{h} = B^2 / \left[4 \pi \sum_s \langle \gamma_s \rangle n_s m_s c^2\right]$, where $\langle \gamma_s \rangle$ is the cell-averaged Lorentz factor and $\rho_s = m_s \, n_s$ the mass density per species, $s \in \{i, e^-, e^+\}$, normalized with the Goldreich-Julian mass density $\rho_\mathrm{GJ} = m_e  \, n_\mathrm{GJ}$.
    Kelvin-Helmholtz vortices are seen to form along the jet-disk interface.
    The inset panels show the density-averaged energy per particle $\langle \xi \rangle = \sum_s \langle \gamma_s \rangle \rho_s c^2 / \sum_s \rho_s c^2$ for different regions of the simulation domain.
    The \textit{green} contours outline the magnetic flux surfaces.
    The \textit{dashed blue} contours denote the ergosphere boundary.
    }
    \label{fig:densmaps}
\end{figure}

\paragraph{Results.---}
The accretion flow dynamic is well summarized in the time series of horizon-penetrating fluxes, where we define the mass accretion rate $\dot{M} = - \iint_A \rho u^r \sqrt{-g} \mathrm{d}\theta\mathrm{d}\phi$ and magnetic flux $\Phi_\mathrm{B} = \tfrac{1}{2} \iint_A |B^r| \sqrt{h} \mathrm{d}\theta\mathrm{d}\phi$ on the horizon ($r=r_\mathrm{h}$).
Here, $\rho$, $u^\mu$, $g$, $h$, and $B^r$ are the mass density, four-velocity, metric determinant, spatial metric determinant, and the radial component of magnetic field as seen by local fiducial observers (FIDOs) that are normal to spatial hypersurfaces, respectively.
The quantity known as the MAD parameter, defined as $\phi_\mathrm{BH} = \Phi_\mathrm{B} / \sqrt{\dot{M} r^2_\mathrm{g} c}$, is known to saturate at ${\sim}50$ \citep{yuan14}.
After it saturates, a flux eruption event occurs, where the magnetic flux is (partially) pushed away from the BH.
This leads to reconnection of the upper and lower hemispheres of the BH magnetosphere in the equatorial plane \citep[][]{ripperda22,davelaar23,vos24timelags}, and a resulting drop in $\Phi_\mathrm{B}$.
We also find that the pairs created by photon-photon annihilation leave an imprint on the system's flux eruptions, as a significant (vacuum) gap can form during the peak of the flux eruption that is rapidly filled by a pair production cascade (Fig.~\ref{fig:densmaps}b).

In Fig.~\ref{fig:timeseries}, we show direct evidence for a quasi-steady axisymmetric MAD accretion state achieved with GRRPIC methods.
The top panel clearly outlines sharp decreases in the mass accretion rate $\dot{M}$, which are direct consequences of the flux eruption events, where the accretion flow is temporarily halted because of a magnetic field barrier created by the flux eruption.
This relates to the sawtooth structure seen for $\Phi_\mathrm{B}$, where the exponential decay is related to the reconnection event (or flux eruption). 
The general shape of growing $\Phi_\mathrm{B}$ with the sawtooth structures of ${\sim}50 \, r_\mathrm{g} / c$ in length combined with significant amplitude changes have not been seen before in other GRPIC simulations \citep[cf.][]{crinquand21,bransgrove21,galishnikova23}. 
The MAD state, corresponding to reaching the limit $\phi_\mathrm{BH} \gtrsim 50$, is established from the very beginning of the simulation.
For $T \gtrsim 650 \, r_\mathrm{g}/c$, we find that both $\Phi_\mathrm{B}$ and $\dot{M}$ start to saturate.

\begin{figure*}
    \centering
    \includegraphics[width=\textwidth]{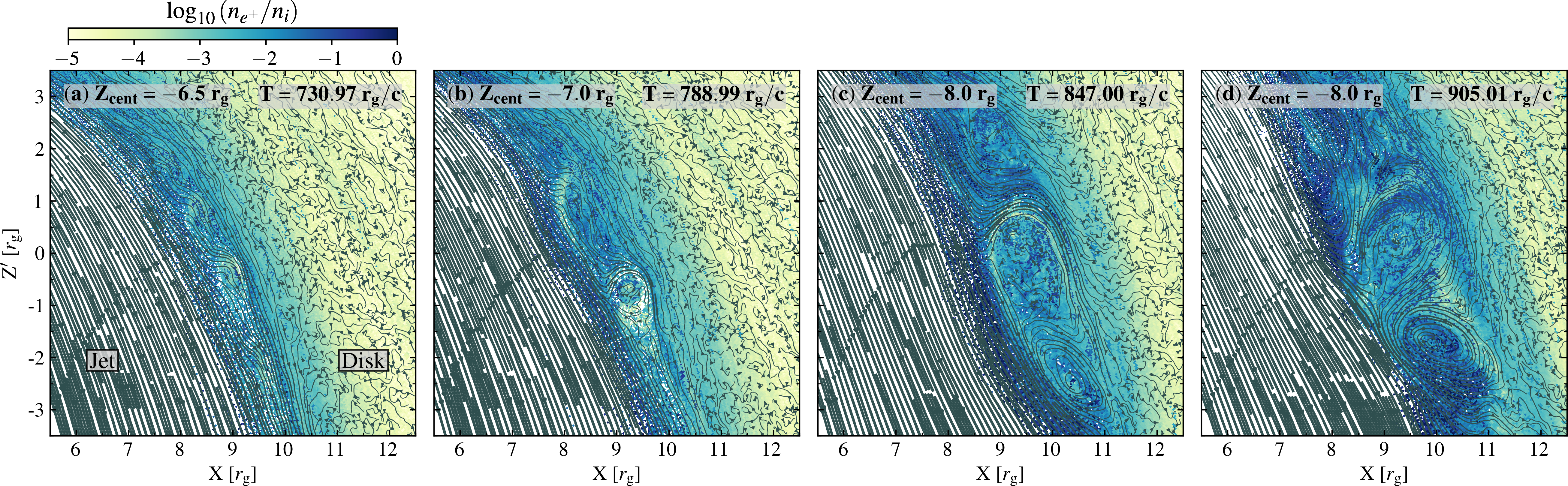}
    \caption{Growth of a Kelvin-Helmholtz-like mode in the jet-disk boundary over time, where mixing between the disk ions ($i$) and the jet positrons ($e^+$) occurs as shown by the number density fraction ($n_{e^+}/n_i$). The swirl-like motion of the KH vortices is clearly denoted by the three-velocity streamlines.
    The streamline thickness is scaled according to the magnitude of the velocity.
    The panels are shown as a function of $\mathrm{Z}^\prime= \mathrm{Z} - \mathrm{Z}_\mathrm{cent}$.
    The shown instances are time-averaged over a window of $\Delta \mathrm{T} \approx 7 \, r_\mathrm{g}/c$.
    }
    \label{fig:mixing}
\end{figure*}

Fig.~\ref{fig:densmaps} shows the characteristics of the MAD accretion state in our collisionless simulation.
Panel \ref{fig:densmaps}a shows a still ongoing flux eruption where the accretion flow is receding and the magnetic field lines are reconnecting. Panel \ref{fig:densmaps}b displays the moment when the accretion flow reinstates itself, creating a clear kink in the equatorial current sheet \citep[a manifestation of the relativistic drift-kink instability;][]{werner21,werner24}.
The disk-magnetosphere system can be divided into several core regions that are of dynamical interest: (i) the inner \textit{jet}, (ii) the jet \textit{sheath} (i.e., the jet-disk interface), (iii) the equatorial \textit{reconnection} region, (iv) the \textit{inner} (polar) magnetosphere (enclosing the spark gap), and (v) the \textit{disk}.
The overdense equatorial plane is in part the result of the enclosing properties of the poloidal magnetic-field loop structure.
Moreover, the tearing modes of the equatorial current sheet instigate the creation of pairs, which further contributes to the overdense equatorial plane region.
These pairs are created in the vicinity of X-points (locations of significant non-ideal electric fields) of the magnetically reconnecting layer, where efficient particle acceleration is known to occur \citep{ripperda20,ripperda22,vos24}.

Fig.~\ref{fig:mixing} highlights that significant velocity shear is found along the jet sheath; 
associated vortices are similar to those seen for the Kelvin-Helmholtz (KH) instability.
The vortices can give rise to magnetic reconnection and subsequent plasmoid formation \citep[shown in a microscopic scenario in][]{sironi21}.
Additionally, it becomes apparent from the $n_{e^+} / n_i$ fraction that significant mixing occurs between matter associated with the jet (i.e., positrons) and matter associated with the disk (i.e., ions), which interface in the jet sheath region. 
However, the ions do not mix effectively into the inner jet region, as is confirmed in Fig.~\ref{fig:mixing}.
We also note that the jet sheath layer is quite magnetized with $\sigma_\mathrm{h} \approx 1$ (Fig.~\ref{fig:densmaps}), indicating that magnetic fields are dynamically important.

We find that the flux eruptions occasionally result in the stretching and curling of magnetic fields along the jet sheath.
This is caused by an initially non- or slowly rotating magnetic field line that is suddenly set into motion after it enters the BH ergosphere creating a torsional Alfv\'en wave. 
The oscillations of the Alfv\'en wave result in sufficient velocity shear to bring about KH-like behavior.
Some of the KH eddies \citep[see also][]{davelaar23} eventually form circular (island) magnetic field structures \citep{sironi16,vos24}.
These structures then grow via the surrounding swirl-like flows which drag magnetic fields that relax into a circular shell around the original structure.

\begin{figure}
    \centering
    \includegraphics[width=0.48\textwidth]{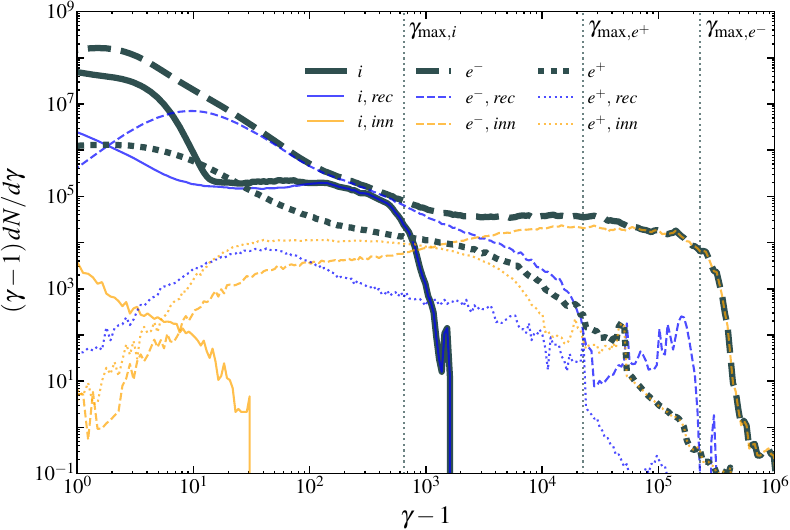}
    \caption{Two most prominent particle acceleration regions correspond to the equatorial reconnection layer (\textit{rec}) and inner polar BH magnetosphere (\textit{inn}).
    Particle energy distribution from distinct accretion flow regions for the ion ($i$, solid), electron ($e^-$, dashed), and positrons ($e^+$, dotted lines). 
    The thick lines correspond to the spectra calculated for all particles in the $r\leq25r_\mathrm{g}$ domain.
    The spectra are averaged over the time window $T \in [726.85,736.14] \, r_\mathrm{g}/c$ corresponding to Fig.~\ref{fig:densmaps}a.
    The vertical lines denote the $\gamma_\mathrm{max}$ values that disregard the tail of the distribution.
    More details of the spectra are outlined in the Sup.~Mat.~\citep{supplemental}.
    }
    \label{fig:spectra}
\end{figure}

Fig.~\ref{fig:timeseries}d shows the evolution of $\gamma_\mathrm{max}$ over time (which is also spatially identifiable with $\langle \xi \rangle$ in Fig.~\ref{fig:densmaps}), where it becomes clear that $\gamma_{\mathrm{max},e^+} \gtrsim \gamma_{\mathrm{max},e^-}$ during selected flux eruption events (i.e., the vertical lines). 
These events coincide with the restoration of the accretion flow after a flux eruption.
This connects to the findings outlined in Fig.~\ref{fig:spectra} \citep[and Sup.~Mat.;][]{supplemental}, where we establish the spatial locations of $\gamma_\mathrm{max}$ per particle species.
These numbers should be evaluated as a fraction of the maximal Lorentz factor implied by the potential drop $\hat{\gamma}_\mathrm{max} = e a B r_\mathrm{g} / m_e c^2 \simeq 2 \times 10^5$ \citep{parfrey19}, taking into account the increase in $B$ over time.
This is verified in Fig.~\ref{fig:timeseries}b where $\Phi_\mathrm{B}$ increases ${\sim}20$ times with respect to the starting value, pushing the simulation closer to a realistic separation of scales.
The inner polar magnetosphere regions produces the highest $\gamma_\mathrm{max} \sim 10^5$ for the electrons and, to a lesser degree, also for the positrons.
The ions plateau around $\gamma_{\mathrm{max},i} \sim 10^3$, but they do reach the same total energy when one compensates for the mass difference as $m_i \gamma_{\mathrm{max},i} / m_e \sim 10^5$.
We note that $\gamma_{\mathrm{max},e^-}$ and $\gamma_{\mathrm{max},e^+}$ can originate in different regions, where the positrons are mainly accelerated in the equatorial reconnection region.
The timing of this acceleration coincides with the restoration of the accretion flow after a reconnection event.
The pairs created near the X-point are collected at the front of the restoring accretion flow.
Occasionally, a considerable (vacuum) gap forms on a similar time-scale, which is rapidly filled with pairs that can then affect and slightly oppose the restoration of the accretion flow (Fig.~\ref{fig:densmaps}b).

The most accelerated electrons originate within and near the ergosphere region where pair creation dynamics associated with spark gaps are known to be important \citep{crinquand20}. 
At the high optical depth to pair creation used here, the gap is small (${\sim}0.01 r_\mathrm{g}$) and pairs are created in intermittent bursts originating close to the BH that are then accelerated outward (i.e., electrons) or inward (i.e., positrons).
Fig.~\ref{fig:densmaps} shows the two distinct channels for the pair production cascade, which can be advected into the equatorial region (mainly accelerating positrons, Fig.~\ref{fig:densmaps}b) or along the polar (jet) regions (mainly accelerating electrons, Fig.~\ref{fig:densmaps}a).
Especially the equatorial pair cascade (associated with significant unscreened electric fields \citep{supplemental}), creating an equatorial evacuated (\textit{bubble}-shaped) structure (Fig.~\ref{fig:densmaps}b), has not been seen before. 
For the modest multiplicity pair plasma (i.e., $n_{e^\pm}/n_\mathrm{GJ} \lesssim 5$), we note that the roles (i.e., inflowing/outflowing) of the leptons are primarily determined by the choice in the sign of $\bm{\Omega}_\mathrm{h} \cdot \bm{B}$; at higher multiplicities, this asymmetry should vanish and polar-cap acceleration should be highly diminished while reconnection-driven particle acceleration should remain mostly unchanged.

\paragraph{Discussion.---}
We have highlighted the complex interplay between accretion-driven behavior, in the form of magnetic flux eruptions, and magnetospheric effects, with the included inverse-Compton scattering and pair production.
We identified the most prominent channels of particle acceleration, which are associated with pair production at the polar caps (pairs only) and with magnetic reconnection in the equatorial current sheet (pairs and ions; Fig.~\ref{fig:spectra}).
Additionally, we find that the jet is loaded with leptons solely created by pair production and virtually no ions are mixed into it (Fig.~\ref{fig:mixing}).
Significant mixing between disk and jet matter occurs at the disk-jet interface, primarily because of prominent KH-like modes that grow over time.
These modes could lead to additional particle acceleration on the jet-disk boundary on larger scales \citep{sironi21}.
The specifics of the gap remain debated \citep{levinson11,levinson17,levinson18,broderick15,hirotani16,crinquand20,chen20}, but from our simulation it becomes clear that the majority of highly energetic pairs are produced there, where two distinct channels can be identified for our modest multiplicities (Fig.~\ref{fig:densmaps}).
Furthermore, the BH spin influences gap characteristics, warranting further investigation.
Future 3D studies will be even more informative on the kinetic effects related to the flux tubes themselves \citep[e.g.,][]{vos24timelags}.

\paragraph{Conclusion.---}
In this study, we presented the first numerical simulation that combines collisionless accretion (ion-electron disk) and magnetospheric (inverse-Compton scattering and pair creation) effects self-consistently using GRRPIC methods.
We highlight the intricate dynamics of a quasi-steady MAD accretion state that is currently preferred for several of the most-observed supermassive BH systems \citep[e.g.,][]{eht1,eht22sgrai}.
The flux eruptions associated with these systems bring about strongly magnetized, largely vacated regions near the central BH, where strong particle-acceleration processes are triggered --- mostly in the equatorial current sheet and inner polar magnetospheric regions.
The cyclic nature of MAD flux eruptions makes them prime candidates for interpreting the periodic flares of high-energy emission observed from these systems \citep[][]{dodds-eden09,gravity20}.

\begin{acknowledgments}
The authors acknowledge informative and constructive discussions with F.~Bacchini, J.~Davelaar, H.~Olivares, and C.~Gammie.
Special thanks is extended to H.~Falcke for his support in acquiring the computational resources.
This project has received funding from the European Research Council (ERC) under the European Union’s Horizon 2020 research and innovation program (Grant Agreement No. 863412). 
JV acknowledges support from the Dutch Research Council (NWO) and SURF which provided the utilized computing resources through grants No. EINF-7054 and NWO-2024.004.
BC acknowledges computing resources that were provided by TGCC under the allocation A0150407669 made available by GENCI.
MM and JV acknowledge support by the NWO grant No.~OCENW.KLEIN.113. KP acknowledges support from the Laboratory Directed Research and Development Program at Princeton Plasma Physics Laboratory, a national laboratory operated by Princeton University for the U.S.\ Department of Energy under Prime Contract No.\ DE-AC02-09CH11466.
\end{acknowledgments}

\bibliographystyle{apsrev4-2}
\bibliography{references_short} 

\appendix

\newpage
\mbox{}
\newpage

\widetext
\begin{center}
\textbf{\large \hspace{1.2cm}Particle acceleration in Collisionless Magnetically Arrested Disks:\newline Supplemental Material}  
\end{center}

\vspace{1em}

\makeatletter

In this Supplementary Material, we provide more details on the initial conditions of the simulation as well as the procedure that underlies the spatially decomposed spectrum calculations.

\subsection{Kerr-Schild coordinate system}
Following generic numerical black hole modeling practice, we adopt horizon-penetrating Kerr-Schild coordinates (in the 3+1 decomposition \citep{misner73} and closely following \citet[][]{komissarov04}), which are defined according to
\begin{align}
    &ds^2 = - \alpha^2 dt^2 + h_{ij} (dx^i + \beta^i dt) (dx^j + \beta^j dt), \hspace{2em} \textrm{with} \\
    &h_{ij} = 
    \begin{pmatrix}
    \Sigma \sin^2 \theta / \rho^2 & -a \sin^2 \theta (1+z) & 0\\
    -a \sin^2 \theta (1+z) & 1 + z & 0\\
    0 & 0 & \rho^2
    \end{pmatrix}. \nonumber 	
\end{align}
Here, $\alpha=1/\sqrt{1+z}$ is the lapse, $\beta = \left(0, z / (1 + z), 0\right)$ is the lapse, and $h_{ij}$ is the spatial metric with $\rho^2 = r^2 + a^2 \cos^2 \theta$, $\Delta = r^2 - 2Mr + a^2$, $\Sigma = (r^2 + a^2)^2 - a^2 \Delta \sin^2 \theta$, and $z = 2Mr / \rho^2$ for $i,j \in \{r, \theta, \phi\}$.
For completeness, we note that $M$ is the black hole mass and that we have adopted geometrized units with $G=c=1$ here.
For the non-geometrized version of the coordinates, one can simply replace $M$ with gravitational radius $r_\mathrm{g}=GM/c^2$.
We note that the familiar form of the metric determinant is recovered as $\sqrt{-g} = \alpha \sqrt{h}$.
All quantities are expressed in Kerr-Schild units throughout the main text and the supplemental material; figures are typically shown as a function of $\mathrm{X,\, Z} = r \cos \theta, \, r \sin \theta$.
Lastly, we note that the underlying computational grid is uniformly spaced in $\hat{\theta}$ and logarithmically spaced in $\hat{r}$.


\subsection{Initial conditions, magnetic field initialization, and parameters exploration}

\begin{figure}[h]
    \centering
    \includegraphics[width=\linewidth]{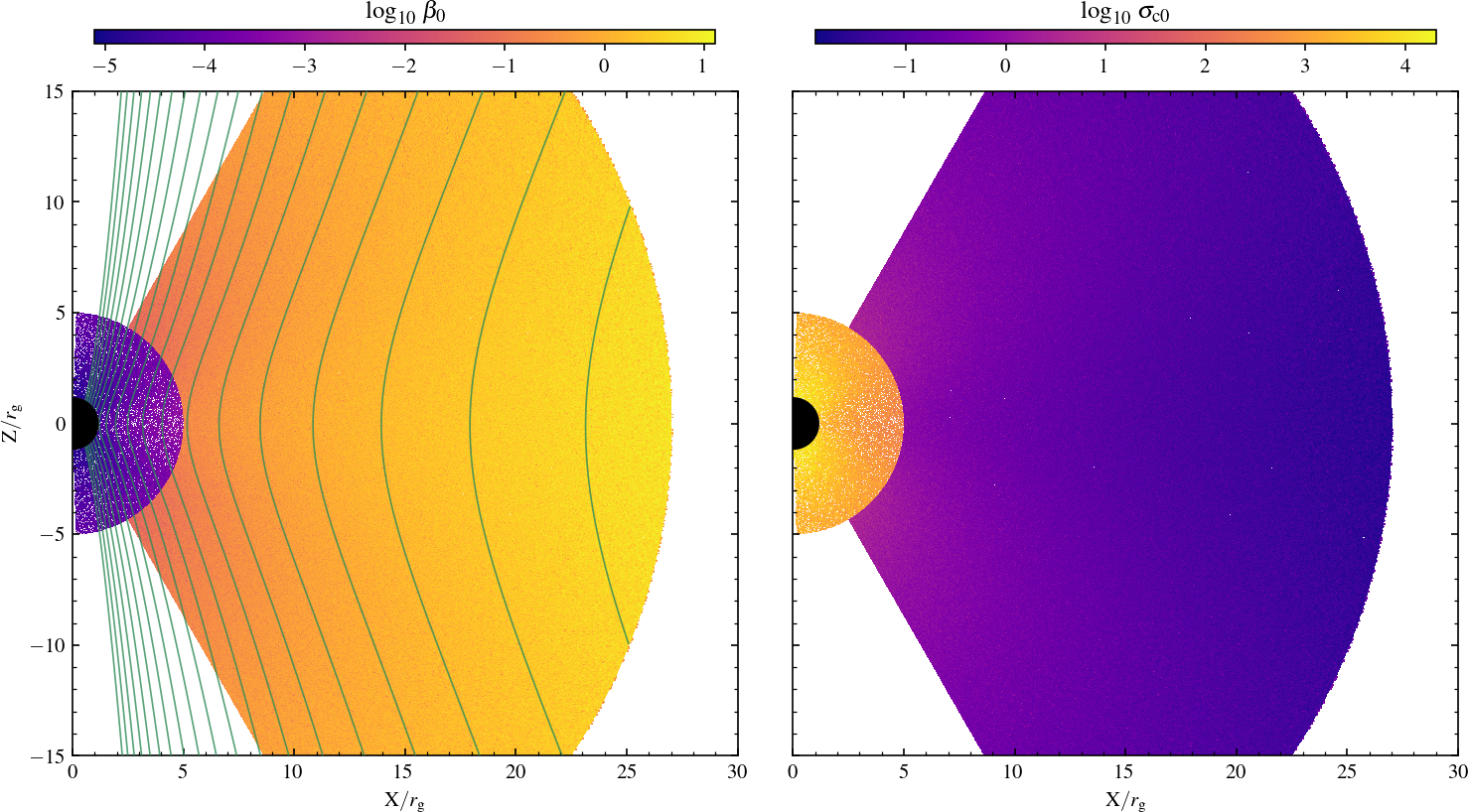}
    \caption{Initial conditions of the accretion setup, for which we display $\beta_0$ and (cold) magnetization $\sigma_{c0}$. 
    The logarithmically-scaled magnetic flux surfaces are shown in \textit{green}. 
    The two distinct regions are the pair-plasma sphere (high $\sigma$, low $\beta$) and the ion-electron disk (low $\sigma_c$, moderate $\beta$).
    }
    \label{fig:initial_conditions}
\end{figure}

Most of the details on the initial particle composition of the accretion flow have already been outlined in the main text, where the accretion environment contains three particle species (i.e., ion-pair plasma) that are organized in a ($8n_\mathrm{GJ}$, ion-electron plasma) wedge combined with a ($1n_\mathrm{GJ}$, electron-positron plasma) seeding sphere ($r < 5 r_\mathrm{g}$).
We explicitly show the initial conditions of $\beta = \sum_s \beta_s$ \citep{fitzpatrick14} and (cold) magnetization $\sigma_\mathrm{c} = B^2 / \left[ 4 \pi \sum_s m_s n_s c^2 \right]$ in Fig. \ref{fig:initial_conditions}.
After an initial phase of strong pair-production (filling the empty space), we find that the accretion flow relaxes into a configuration dictated by the initialized magnetic field over $100$ \rgc.

We initialize the magnetic fields ($\bm{B} = \nabla \times \bm{A}$) via the following vector potential $\bm{A} = (0, 0, A_\phi)$ where
\begin{equation}
    A_\phi = B_0 \left( \tfrac{r}{r_\mathrm{g}} \right) e^{-(r / r_\mathrm{g}) / 400} \sin^2 \theta,
\end{equation}
which is inspired by what is typically used to create a magnetically arrested disk (MAD) accretion state in GRMHD simulations \citep[e.g.,][]{vos24timelags}.
Customarily, however, the GRMHD magnetic field configuration is partly determined by the initialized hydrostatic disk solution \citep[e.g.,][]{fishbone76}, which is technically not applicable for GRPIC methods so the magnetic field configuration inherently resembles a poloidal loop structure.
The field was chosen so that $\bm{\Omega}_\mathrm{h} \cdot \bm{B} > 0$.
We ran the presented simulation for $1000$ \rgc and eventually reached a quasi-steady accretion state with multiple flux eruption cycles \footnote{For clarity, we further nuance what is meant by quasi-steady accretion state. MAD accretion states are characterized by magnetic flux eruption cycles that can strongly perturb the inflow of matter. This effect is reinforced in the relatively confining characteristics of axisymmetric simulations \citep[e.g.,][]{proga03}. So, when we speak of a \textit{quasi-steady} state, we refer to the flattening over time of the $\dot{M}$ and $\Phi_\mathrm{B}$ timeseries shown in Fig.~1 of the main text.}.  

For completeness, we note that while the contents of the main text and supplemental material are based on a single simulation, many exploratory runs have been undertaken. 
The most comparable simulations were performed with a lower ion mass ($m_i = 50 m_e$) and variations in initialized disk number densities ($8 n_\mathrm{GJ}$ and $12 n_\mathrm{GJ}$).
The general findings for those simulations correspond well to what was reported for the simulation featured in the main text and supplemental material.
These other simulations also show the formation of prominent Kelvin-Helmholtz (KH) eddies, as well as periodically repeating flux eruptions.
Furthermore, most of the exploratory runs were conducted with a significantly smaller resolution ($N_r \times N_\theta = 864^2$) for which we found similar behavior (albeit less well resolved).

\subsection{Pair production mechanics and separation of scales}
In our simulation the $\gamma$-ray photons are evolved explicitly as a neutral fourth particle species, which comes about when a time-independent, uniform, and monochromatic soft background photon inverse Compton scatters of a relativistic lepton. 
The $\gamma$-rays are then able to pair-produce with the background radiation field.
The details of the Monte Carlo procedure that lies at the basis of this can be found in \citet{levinson18} and \citet{crinquand20}.
These dynamics are largely determined by the choice in fiducial optical depth $\tau_0 = n_0 \sigma_\mathrm{T} r_\mathrm{g} = 50$, where $\sigma_\mathrm{T}$ is the Thompson cross-section and $n_0$ is the fiducial number density, which indicates that we operate in a relatively high optical depth regime that is likely significantly higher than is assumed for low-luminosity AGN (such as Sagittarius A$^\ast$ and M87$^\ast$).
The other fiducial parameters are the normalized magnetic field, $\tilde{B}_0 = e B_0 r_\mathrm{g} / m_e c^2 = 10^4$, where the initial magnetic field strength is $B_0 = 10^4 m_e c^2 / |e|$ (see also the extended discussion in \citet{parfrey19}).
The normalized energy of the background photons, $\tilde{\epsilon}_0 = \epsilon_0 / m_e c^2 = 10^{-2}$, so that $\tilde{B}_0 \tilde{\epsilon}_0 = 100$ at initialization (compared to $\tilde{B}_0 \tilde{\epsilon}_0 \sim 10^6$ for, e.g., Sagittarius A$^\ast$). 
This reflects an artificial shrinkage in the separation of scales, which is required to make the numerical study feasible \citep[][]{crinquand20}.
Nevertheless, as becomes clear from Fig.~1 in the main text, the magnetic flux and corresponding magnetic field strength increase by a factor $20$ over the duration of the simulation, which indicates movement towards a more realistic separation of scales. 

We generally resolve the relativistic (and non-relativitic) electron skin depth $d_{e} = \sqrt{\langle \gamma_{e} \rangle m_e c^2/ 4 \pi n_\mathrm{GJ} e^2} \sim r_\mathrm{g}/\sqrt{\tilde{B}_0}$ at a resolution of $N_r \times N_\theta = 1440^2$ over the evaluated time, which implies that the ion skin depth $d_i = \sqrt{\langle \gamma_{i} \rangle m_i / \langle \gamma_{e} \rangle m_e} \, d_e$ is also well resolved.
The reported simulation has a significant dynamical range in mass density and magnetic field strength (see Fig.~\ref{fig:initial_conditions} and Fig.~2 of the main text). 
During the quasi-steady accretion stages ($T \gtrsim 650 \, r_\mathrm{g}/c$), particle phase space is well sampled, that is, the average number of electrons and positrons per cell is of O(60).
The total number of particles in the simulation is typically ${\sim}3 \times 10^8$, consisting primarily of electrons and positrons.

\subsection{Spatially decomposed spectra calculations} \label{app:mask}

\begin{figure}
    \centering
    \includegraphics[width=\textwidth]{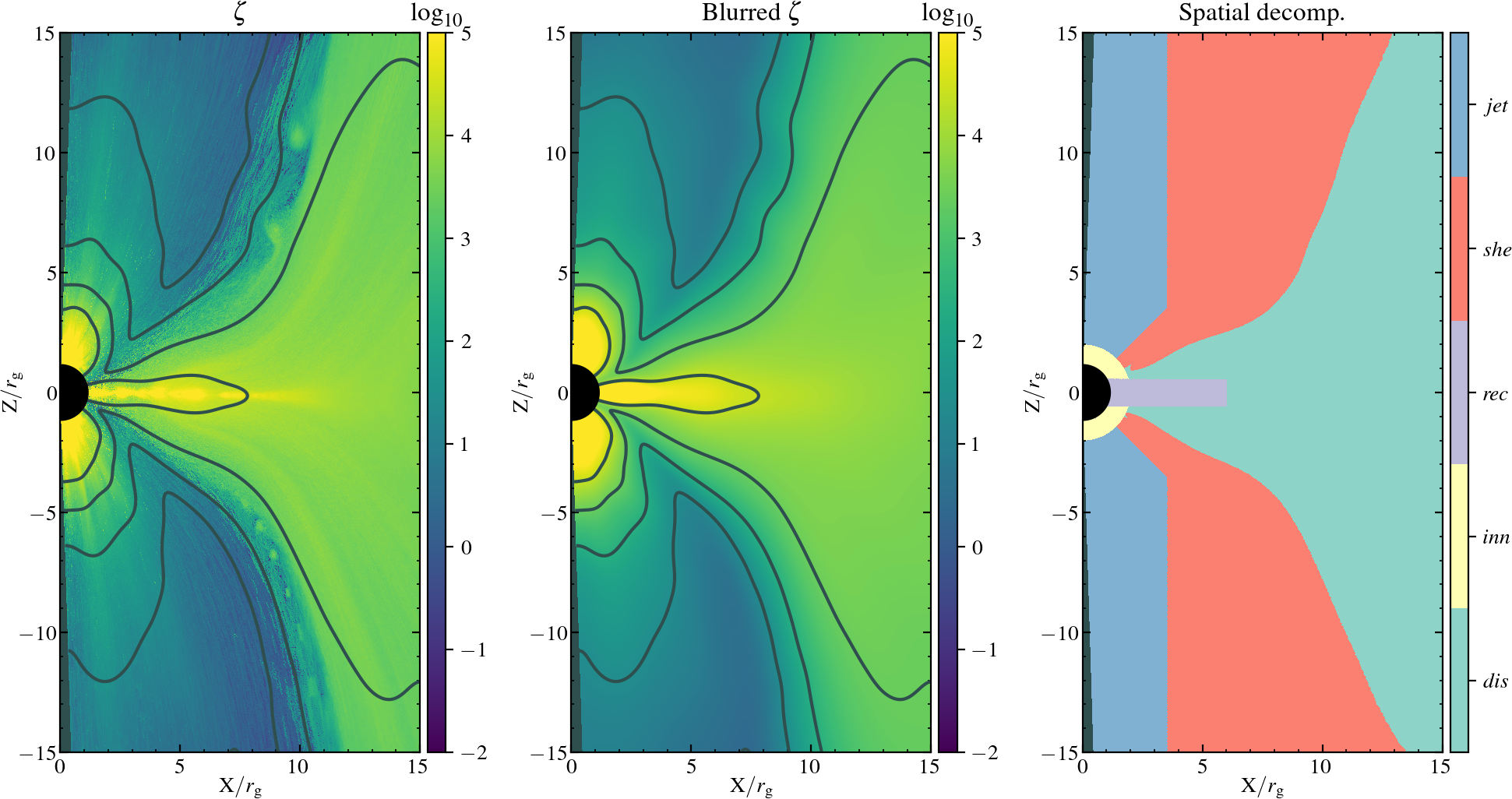}
    \caption{Schematic showing how the spatially-dependent spectra are calculated. The quantity we evaluate is the normalized energy density $\zeta = \sum_s \langle \gamma_s \rangle \rho_s c^2 / \rho_\mathrm{GJ} c^2$ with $\rho_\mathrm{GJ} = m_e n_\mathrm{GJ}$ and $\rho_s = m_s n_s$ for all particle species $s \in \{e^-, e^+, i\}$. This quantity is time-averaged over a period $T \in [726.85,730.33] \, r_\mathrm{g}/c$ and subsequently blurred to obtain a smooth dissection between the disk (\textit{dis}) and jet sheath (\textit{she}) regions. Based on the empirically found criterion (such as $\zeta > 1950$), one then makes the distinction (for the disk region). 
    The remaining spatially decomposed regions are the jet (\textit{jet}), equatorial reconnection (\textit{rec}), and inner (polar) magnetosphere (\textit{inn}) regions.
    The contours illustrate logarithmically scaled variations in the blurred $\zeta$ quantity. 
    }
    \label{fig:maskapp}
\end{figure}

As discussed in the main text, we differentiate five distinct regions in which we evaluate the particle energy spectra; these regions are denoted as jet (\textit{jet}), jet sheath or jet-disk interface (\textit{she}), equatorial reconnection (\textit{rec}), inner (polar) magnetosphere (\textit{inn}), and disk (\textit{dis}) regions.
The decomposition of these regions is shown in Fig.~\ref{fig:maskapp}.
The \textit{mask} is created with geometrical and dynamic criteria. 
For simplicity, we set the speed of light $c=1$ in the remainder of this section.

Effectively, only the boundary between the disk and the jet sheath is determined by a dynamical criterion; 
when the quantity $\zeta > 1950$ (Fig.~\ref{fig:maskapp}), we define it as disk, while when this criterion is not satisfied, we assign it as jet sheath.
Here, we define $\rho \langle \gamma \rangle = \sum_s m_s n_s \langle \gamma_s \rangle$, which contains the contribution of all species $s \in \{i, e^-, e^+\}$.
The remaining areas are the jet region (conical, [$\theta < \pi/4 \vee \theta > 3\pi/4$] \& $r < 5 r_\mathrm{g}$, and then cylindrical, $r > 5 r_\mathrm{g}$ \& $\mathrm{X} < 5\sin[\pi/4]$), the reconnection region ($\mathrm{X} < 6r_\mathrm{g}$ \& $|\mathrm{Y}| < 0.5r_\mathrm{h}$), and the inner magnetosphere ($r_\mathrm{h} < r < 2 r_\mathrm{g} \setminus \mathrm{\textit{rec}}$), which can be obtained from the figure itself.
The geometrical areas are leading over the dynamically assigned areas.
We note that the spectra are only evaluated for $r \in [r_\mathrm{h}, 25 r_\mathrm{g}]$.

The spectra are created from reprocessed data of the original simulation that are outputted at a higher cadence and then put through the spatial-separation pipeline outlined in the previous paragraph. 
These individual spectra are then averaged to reduce the low-amplitude noise.
The calculation of these spectra is a relatively costly endeavor. 
We verified that the resulting spectra remain consistent under (small) variations in the spatial decomposition mask.

Here, we comment on how $\gamma_\mathrm{max}$ is determined.
This is done by finding the Lorentz factor corresponding to the 80th percentile of the cumulative distribution function of the function $f_s =\gamma_s^{p} dN/d\gamma_s$, where $dN/d\gamma_s$ is the particle energy spectrum and $p$ the empirical power law index that differs per particle species $s$. 
For electrons and positrons, we use $p_{e^\pm} = 1.5$.
As the ions have a comparatively larger thermal component, we need to adopt a larger power law index to be most accurate, for which we choose $p_{i} = 2.5$.

\begin{figure}
    \centering
    \includegraphics[width=0.48\linewidth]{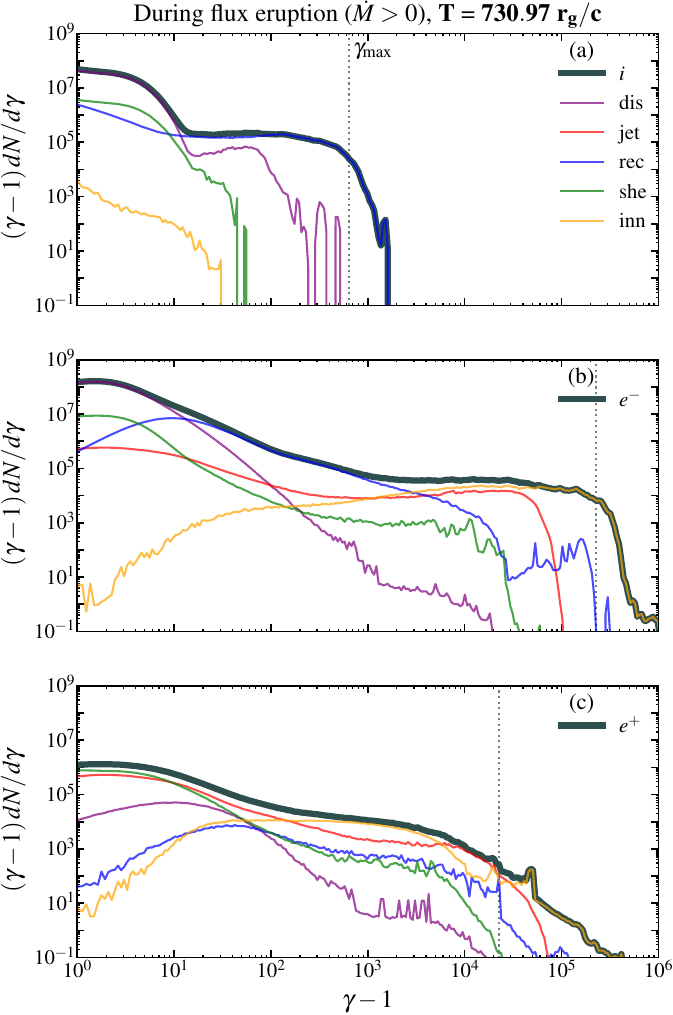}
    \hfill
    \includegraphics[width=0.48\linewidth]{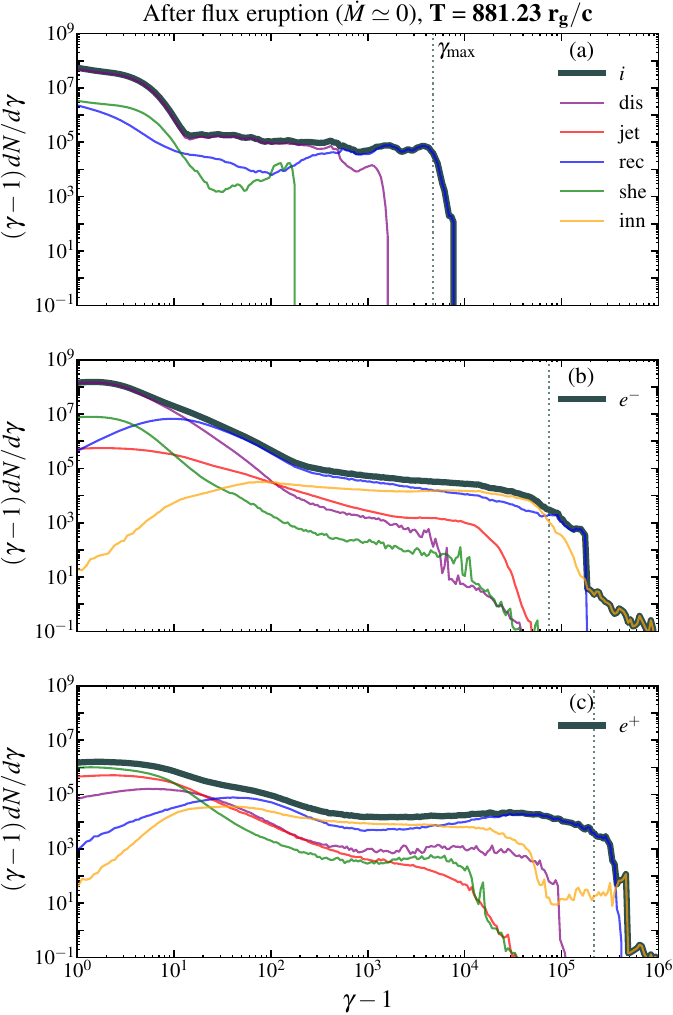}
    \caption{
    Particle energy distributions showing the regions where the particle species are primarily accelerated --- strongest acceleration takes place in the reconnection and inner magnetospheric regions.
    The thick lines indicate the combined contributions of all regions for the ion ($i$), electron ($e^-$), and positrons ($e^+$) in separate vertical panels. 
    Five distinction regions are identified as disk (\textit{dis}), inner jet (\textit{jet}), equatorial reconnection layer (\textit{rec}), inner black hole magnetosphere (\textit{inn}), and the jet sheath (\textit{she}).
    The left and right spectra are averaged over the time windows $T \in [726.85,736.14] \, r_\mathrm{g}/c$ and $T \in [874.45, \, 887.50] \, r_\mathrm{g}/c$, respectively, and correspond to the top and bottom panels of Fig.~2 of the main text.
    The vertical lines denotes the $\gamma_\mathrm{max}$ values, where the tail of the distribution is not considered.
    The mask used to create the spatially-dependent spectra is outlined in Fig.~\ref{fig:maskapp}.
    }
    \label{fig:spectra_supplemental}
\end{figure}

For completeness, we show all the components of the mapped spectra for the accretion states shown in the top and bottom panels of Fig.~2 in the main text.
These panels correspond to the time instances $\mathrm{T} = 730.97 \, r_\mathrm{g}/c$ and $\mathrm{T} = 881.23 \, r_\mathrm{g}/c$, and are averaged over time windows $T \in [726.85,736.14] \, r_\mathrm{g}/c$ and $\mathrm{T} \in [874.45, \, 887.50 ] \, r_\mathrm{g} / c$.
For the state corresponding to $\mathrm{T} = 881.23 \, r_\mathrm{g}/c$, we find that $\gamma_{\mathrm{max},e^+} \gtrsim \gamma_{\mathrm{max},e^-}$, which coincides with a strong pair creation cascade that commences in the equatorial plane (forming a \textit{bubble} structure, seen clearly in the right panels of Fig.~\ref{fig:appgap}).
Overall, we find that the creation of the associated gap occurs when the magnetic flux eruption is at its peak and the accretion flow has maximally retracted.
The restoration of the accretion flow coincides with a kink mode (as discussed in the main text) where the strongest acceleration occurs.
We note that the equatorial pair cascade (during which $\gamma_{\mathrm{max},e^+} \gtrsim \gamma_{\mathrm{max},e^-}$) is only a relatively rare occurrence and seems contingent on the strength and duration of the preceding flux eruption.
We show the complete spectra of Fig.~4 of the main text in the left panels of Fig.~\ref{fig:spectra_supplemental} that correspond to a general state where $\gamma_{\mathrm{max},i} < \gamma_{\mathrm{max},e^+} < \gamma_{\mathrm{max},e^-}$.
During these periods, the strongest acceleration occurs (for $e^-$) in the polar magnetospheric regions.
The jet and jet sheath regions harbor less energetic electrons and positrons.
For the ions, we see that moderate acceleration occurs in the sheath, which is connected to the regions containing KH-like modes.
The strongest acceleration of ions is still associated with reconnection in the equatorial current sheet region.

Occasionally, we find that the spectra, mostly for $e^-$, display a three-bumped structure, where the first corresponds to the (thermal) disk component, the second originates in the reconnection region, and the third corresponds to the inner magnetospheric particle-creation dynamics, going from low to high Lorentz factor $\gamma$, respectively.
Interestingly, the secondary reconnection-related bump is most intermittent; a late stage of the decay of this component is still visible in the right panels of Fig.~\ref{fig:spectra_supplemental}.
The intermittency of this reconnection component is expected, as the equatorial reconnection activity varies rapidly and is most directly affected by the MAD flux eruption dynamics. 

To assess the statistical significance of the spectra presented in the main text and this supplemental material, we report that the spectra (shown in Fig.~\ref{fig:spectra_supplemental}) contain a total number of $n_\mathrm{tot} \approx 1.5 \times 10^8$ particles.
These are distributed among electrons, positrons, and ions with typical fractions of $f_{e^-} = 0.5$, $f_{e^+} = 0.4$, and $f_i=0.1$, respectively.
The individual spectra are then combined and averaged over 160 time instances, evenly spread over the indicated time window, to mitigate the largest temporal variations. 
These values are most representative for the later stages of the simulation, of which we show the spectra.
The ratios may seem skewed but we emphasize that not all particles have the same ``numerical weight.'' 
The disk-initialized electrons and ions represent a particle packet equivalent to 8 individual particles.

\subsection{Inner light surface and gap dynamics}
As previously established in \citet{crinquand20}, an intermittent gap can open near the inner light surface under certain circumstances.
Following elaborate discussion by \citet{komissarov04}, the light surfaces correspond to the separating surfaces (i.e., $f(\Omega_F, r, \theta) = 0$) of the function
\begin{equation}
    f(\Omega_F, r, \theta) = g_{\phi\phi} \Omega_F^2 + 2 g_{t\phi} \Omega_F + g_{t\phi},
\end{equation}
with field orbital velocity $\Omega_F = - E_\theta / \sqrt{h} B^r$.
Here, $h$ is the determinant of the spatial part of the metric and $g_{\phi\phi}, g_{t\phi}, g_{t\phi}$ are components of the Kerr metric \citep[as outlined in detail in][]{komissarov04}.
We note that the magnetic field configuration used here is new, but we find good correspondence to what was seen before with regard to the spark gap dynamics for purely magnetospheric studies \citep{parfrey19,crinquand20}, except that some of the gap dynamics are largely dictated by the accretion component in our simulation.
For instance, we see that the reinstating accretion flow also leaves an imprint on the inner light surface \citep[][]{crinquand20}, where the otherwise smooth surface extrudes along the equatorial plane.

\begin{figure}
    \centering
    \includegraphics[width=\textwidth]{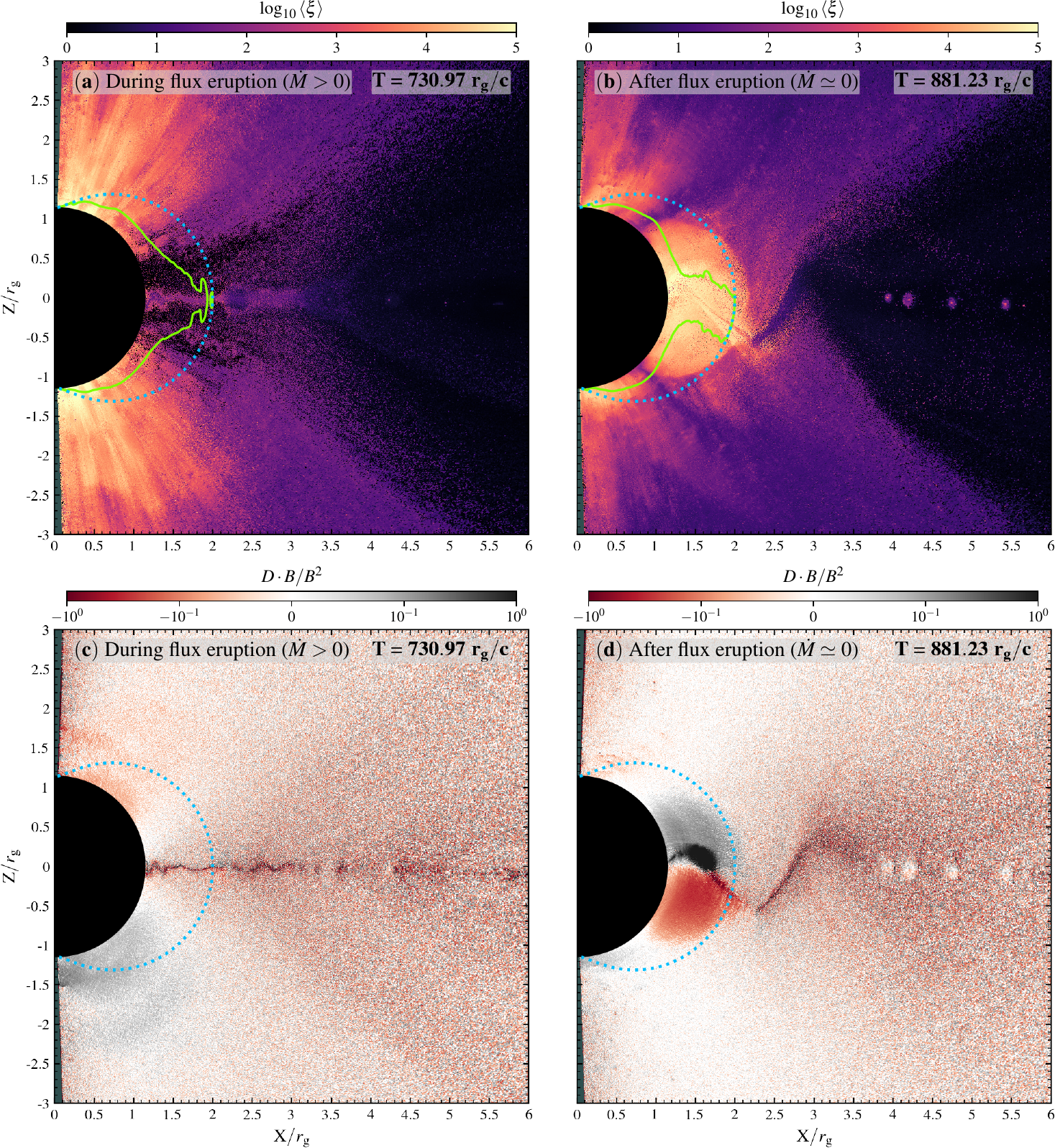}
    \caption{
    Two distinct particle acceleration channels during a flux eruption (left) showing most activity around the poles, which is followed by the recovering of the flow (right) that encounters an equatorial largely evacuated (\textit{bubble}) feature.
    The density-averaged energy per particle $\langle \xi \rangle = \sum_s \langle \gamma_s \rangle \rho_s c^2 / \sum_s \rho_s c^2$ for particle species $s \in \{e^-,e^+,i\}$ shown in panels (a) and (b). 
    These figures correspond to the right sides of the inset panels shown in Fig.~2 of the main text.
    The \textit{green} contours correspond to the \textit{inner light surface} that is determined by the magnetic field orbital velocity $\Omega_F = - E_\theta / \sqrt{h} B^r$, where $h$ is the determinant of the spatial part of the metric.
    The inner light surface is rapidly varying in the equatorial region, so to show the global characteristic we display a contour acquired from a smoothed light surface function.
    The ergosphere is denoted with the \textit{dashed blue} contours.
    The bottom panels (c) and (d) show the amount of unscreened electric field by means of quantity $\bm{D} \cdot \bm{B} / B^2$, where $\bm{D}$ and $\bm{B}$ are the FIDO-measured electric and magnetic fields.
    The colors are log-scaled and simultaneously conserve the sign of the quantity. 
    }
    \label{fig:appgap}
\end{figure}

In the next paragraphs, we further discuss the gap dynamics originally shown in Fig.~2 of the main text and display this region more explicitly in Fig.~\ref{fig:appgap}.
As mentioned in the main text, much remains unclear about the formation, spatial extent, and location of the spark or electrostatic gap, which is characterized by significant quantities of unscreened electric fields denoted by $\bm{D} \cdot \bm{B} / B^2$ in Fig.~\ref{fig:appgap}, where $\bm{D}$ and $\bm{B}$ are the electric and magnetic fields as seen by fiducial observers 
\citep[FIDOs;][]{komissarov04,parfrey19}. 
In Fig.~\ref{fig:appgap}, we show the two distinct pair production channels found in our simulation, which are (i) located in the polar magnetospheric regions and (ii) along the equatorial plane (creating a \textit{bubble} structure, as commented on in the main text).
For completeness, we also show the inner light surface, which could be informative about the location of gap separation surface \citep[between electron and positron;][]{crinquand20}.
However, as discussed before \citep[in, e.g.,][]{levinson18,chen18,chen20,crinquand20}, the gap is thought to be located close to the horizon at high optical depth \citep[$\tau_0 \geq 30$;][]{crinquand20}.
Scenario (i) is consistent with the magnetospheric dynamics of a quasi-periodic, narrow gap opening near the polar axis. 
In Fig.~\ref{fig:appgap}(c), we show the unscreened electric field strength of this state, which indicates that $|\bm{D} \cdot \bm{B}| / B^2 \lesssim 10^{-1}$ near the gap, where the upper and lower hemisphere have an opposite polarity.
In Fig.~\ref{fig:appgap}(b), showing scenario (ii), we find that a \textit{bubble}-shaped equatorial region with highly energetic leptons forms, which is created because of the retracting accretion flow and a strong pair cascade that commences in the newly formed and large vacuum gap.
As seen in Fig.~\ref{fig:appgap}(d), the associated unscreened electric field is large with $|\bm{D} \cdot \bm{B}| / B^2 \sim 1$, which explains the strong pair production cascade that commences in a relatively short time span.
This type of equatorial pair cascade has not been seen before and is strongly related to the MAD flux eruption dynamics.
It also corresponds to the moment when $\gamma_{\mathrm{max},e^+} \gtrsim \gamma_{\mathrm{max},e^-}$ and highly-energetic positrons become more abundant.
The equatorial relativistic drift-kink mode corresponds to the restoring accretion flow and is also the separatrix between positive and negative unscreened electric fields.

\end{document}